\documentclass[twocolumn,showpacs,preprintnumbers,amsmath,amssymb]{revtex4}

\usepackage{graphicx}
\usepackage{dcolumn}
\usepackage{bm}

\begin{document}


\title{Shaping quantum pulses of light via coherent atomic memory}

\author{M.~D.~Eisaman,$^{1}$ L.~Childress,$^{1}$ A.~Andr\'{e},$^{1}$ F.~Massou,$^{1}$ A.~S.~Zibrov,$^{1,2,3}$ and M.~D.~Lukin$^{1}$}
\affiliation{$^{1}$Physics Department, Harvard University, Cambridge, MA 02138, USA \\
$^{2}$Harvard-Smithsonian Center for Astrophysics, Cambridge, MA 02138, USA \\
$^{3}$P. N. Lebedev Institute of Physics, Moscow, 117924, Russia}



\date{\today}

\begin{abstract}
We describe a technique for generating pulses of light with controllable
photon numbers, propagation direction, timing, and pulse shapes.  The technique is based on preparation 
of an atomic ensemble in a state with a desired number of
atomic spin excitations, which is later converted into a photon 
pulse. Spatio-temporal control over the pulses is obtained by exploiting long-lived coherent memory for photon states and electromagnetically induced transparency (EIT) in an optically dense atomic medium.  Using photon counting experiments we observe 
generation and shaping of few-photon sub-Poissonian light pulses.
We discuss prospects for controlled generation of high-purity n-photon Fock states using this technique.
\end{abstract}

\maketitle


In recent years much effort has been directed toward generating quantum-mechanical states of the electromagnetic field with a well-defined number of light quanta (i.e., photon-number or Fock states).   In addition to being of fundamental interest, these states represent an essential resource for the practical implementation of many ideas from quantum information science such as quantum communication~\cite{qietal}.  Over the past decade, tremendous progress has been made in generating single-photon states by using photon pairs in parametric down-converters~\cite{hong86}, 
single emitters~\cite{michler2000etal, santori02etal}, 
and single atoms in high-finesse cavities~\cite{rempe02, mckeever04etal}.  While parametric down-conversion techniques have recently been used to generate multi-photon states~\cite{waks04etal}, it remains experimentally challenging to implement schemes that allow for simultaneous control over both photon number and spatio-temporal properties of the pulse.  

In this Letter we describe 
a novel technique for generating pulses of light with controllable, well-defined photon numbers, propagation direction, timing, and pulse shapes by exploiting long-lived coherent memory for photon states in an optically dense atomic medium~\cite{lukincolloq03}.  
We experimentally demonstrate key elements of this technique in photon-counting experiments.
This approach combines different aspects of earlier studies on ``light storage"~\cite{liu01etal,phillips01etal} and Raman preparation and retrieval of atomic excitations~\cite{vdwal03etal, kuz03etal, chou04etal}.
It is particularly important in the contexts of long-distance quantum communication~\cite{duan01etal}, and EIT-based quantum nonlinear optics~\cite{schmidt96, matsko99etal, braje03etal}.

In  our approach we first optically pump a large ensemble of $N$ atoms with a three-state ``lambda" configuration of atomic states (see Fig.1a)  in the ground state $|g \rangle$.   Spontaneous Raman scattering~\cite{raymer96} is induced by a weak, off-resonant laser beam 
with Rabi frequency $\Omega_{W}$ and detuning $\Delta_{W}$, referred to as the write laser. 
This two-photon process  flips an atomic spin into the metastable state $|s\rangle$ while producing a correlated frequency-shifted photon (a so-called Stokes photon).
Energy and momentum conservation ensure that for each 
Stokes photon emitted in certain direction there exists exactly one flipped spin quantum in a well-defined spin-wave mode. 
The number of spin wave quanta and the number of photons in the Stokes field thus exhibit strong correlations,  
 analogous to the correlations between 
photons emitted in parametric down conversion~\cite{raymer92etal}. 
As a result, 
measurement of the Stokes photon number $n_S$ 
ideally projects the spin-wave into a nonclassical collective state with $n_S$ 
 spin quanta~\cite{duan01etal}.  
After a controllable delay time $\tau_{d}$ (see Fig.~1b),  the stored spin-wave 
can be coherently converted into a light pulse 
by applying a second near-resonant laser beam with Rabi fequency $\Omega_R$ (retrieve laser), see Fig.~1a.  The physical mechanism for this retrieval process involves EIT~\cite{harris-phystoday, scullybook, matsko99etal, boller, fleischhauer00} and is identical to that employed in previous experiments~\cite{liu01etal, phillips01etal}.  The direction, delay time $\tau_{d}$, and rate of retrieval 
are determined by the direction, timing, and intensity of the retrieve laser, allowing control over 
the spatio-temporal properties of the retrieved pulse (referred to as the anti-Stokes pulse).  
Since the storage and retrieval processes ideally result in identical photon numbers in the Stokes and anti-Stokes pulses~\cite{matsko99etal}, this technique should allow preparation of an $n$-photon Fock state  in the anti-Stokes pulse conditioned on detection of $n$ Stokes photons.

\begin{figure}
\includegraphics[width=7.2cm]{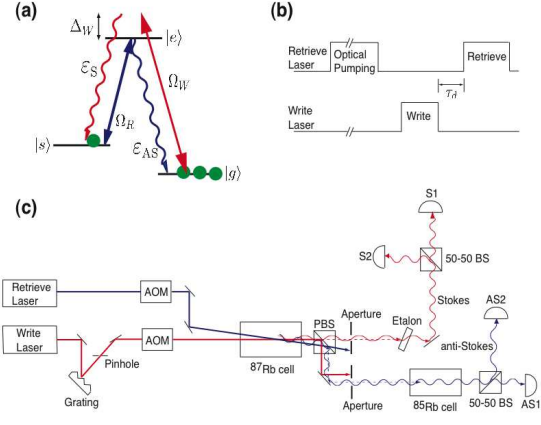}
\caption{\label{fig1} (Color Online) Experimental procedure and apparatus.  ({\bf a}) $^{87}\mbox{Rb}$ levels used in
the experiments ($\mbox{D}_1$ line):  $|g \rangle=|5^{2}S_{1/2},F=1 \rangle$, $|s \rangle=|5^{2}S_{1/2},F=2 \rangle$, and $|e\rangle$ corresponds to $|5^{2}P_{1/2},F'=1 \rangle$ and $|5^{2}P_{1/2},F'=2 \rangle$.
The write (retrieve) laser and Stokes (anti-Stokes) beam are illustrated in red (blue).  The write (retrieve) laser has a waist of 100 $\mu \mbox{m}$ (1 mm).  $\Delta_{W} \approx 1 \mbox{GHz}$.  ({\bf b})  After the optical pumping pulse (provided by the retrieve laser), the $1.6 \, \mu s$-long write pulse is followed by the retrieve pulse after a controllable delay $\tau_{d}$.  ({\bf c}) Schematic
of the experimental setup.  The write and retrieve lasers overlap at a small angle ($\sim 10$ mrad) inside a magnetically-shielded $^{87}\mbox{Rb}$ vapor cell held at $\approx 75^{o}C$.  AOM is acousto-optical modulator.  PBS is polarizing beamsplitter. S1,S2 (AS1,AS2) are avalanche photodetectors (APDs) for the Stokes (anti-Stokes) channel.}
\end{figure}


The experimental apparatus (see Fig. 1c) is similar to that used in our earlier work~\cite{vdwal03etal}.  
The primary experimental challenge lies in transmitting  
the few-photon Stokes and anti-Stokes pulses while simultaneously blocking the write and retrieve laser beams.   A polarizing beamsplitter separates the write (retrieve) laser from the Stokes (anti-Stokes) Raman light, and further filtering is provided by an etalon or an optically-pumped $^{87}\mbox{Rb}$ cell in the write channel, and a $^{85}\mbox{Rb}$ cell in the read channel; this combined filtering separates the write (retrieve) laser from the Stokes (anti-Stokes) Raman light to one part in $10^{9}$ ($10^{12}$).   
Experimentally, we take advantage of the long coherence time  of the atomic memory ($\sim 3 \, \mu$s in the present experiment, see Fig.~3) to create few-photon pulses with long coherence lengths ($\sim \mbox{few} \, \mu \mbox{s}$) that significantly exceed  the time-resolution of the APDs ($\sim 50 \, \mbox{ns}$). This allows us to directly count the photon number in each of the pulses and to directly measure the pulse shapes by averaging the time-resolved APD output over many experimental runs.

Fig.~2a shows  the average number of detected Stokes photons per unit time (photon flux) in the write channel as a function of time during the 1.6 $\mu s$-long write pulse. The magnitude of the photon flux (and
hence the total number of photons in the pulse) is controlled by
varying the excitation intensity.  
The shape of the Stokes pulse changes qualitatively as the total number 
of photons in the pulse exceeds unity:  for pulses containing on average one photon or less, the flux is 
constant in time (more generally it follows the shape of the write laser), whereas for pulses containing more than one photon, the flux increases with time. 

\begin{figure}
\includegraphics[width=7.2cm]{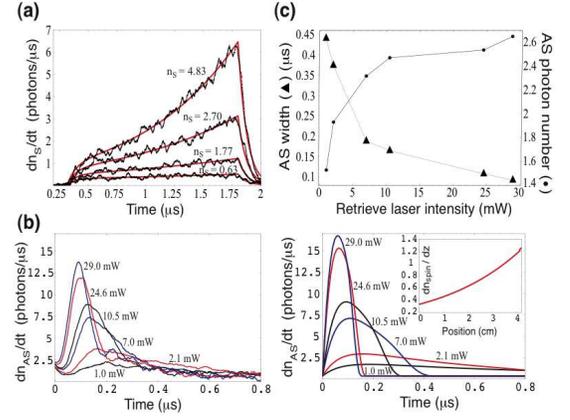}
\caption{\label{fig2} (Color Online) Stokes and anti-Stokes pulse shapes. ({\bf a})  Experimentally measured and theoretically calculated values of the Stokes photon flux $dn_{S}/dt$.
For each plot, $n_{S} = \int dt \, dn_{S}/dt$ represents the total number of photons emitted from the cell.  Write laser power was varied from $25 \, \mu \mbox{W}$ to $100 \, \mu \mbox{W}$.
({\bf b})   Experimentally measured and theoretically calculated values of the anti-Stokes photon flux $dn_{AS} / dt$.
The experimental pulse shapes correspond to a Stokes pulse with $n_S \approx 3$ photons, and the theoretical curves assume an initial spin wave with $n_{spin} =3$ excitations and an optical depth of $\simeq 20$.
Each curve is labeled with the power of the retrieve laser.
({\bf b, inset} ) Theoretical calculation of the number of flipped spins per unit length $dn_{spin}/dt$ (cm$^{-1}$) 
for $n_{spin} = 3$.  
({\bf c}) Measured anti-Stokes pulse width (full-width at half-max) and total photon number as a function the retrieve laser intensity.  
 }
\end{figure}

The observed evolution of the Stokes pulses can be understood qualitatively by considering the mutual 
growth of the photon field and spin excitation: the first flipped spin  
{\it stimulates} subsequent spin excitations which are accompanied by 
increased probability of Stokes photon emission.  This process is governed by  the collective Raman scattering rate $\xi =   \eta |\Omega_W|^2 \gamma / \Delta_W^2 $, which is equal to the product of the optical depth $\eta$ and the single atom scattering rate $|\Omega_W|^2 \gamma / \Delta_W^2 $, where $\gamma$ is the decay rate of $| e \rangle$.   For short excitation times t, we consider the evolution of several Hermite-Gaussian modes of Stokes radiation~\cite{raymer96}, and find that the photon flux is given by $dn_s/dt = \bar{N} \xi (1 + \xi t + ...)$, where $\bar{N}$ is the effective number of transverse modes (from experimental measurements we infer $\bar{N} \sim 4$).  This prediction for $dn_s/dt$  agrees with our observation that the flux will increase with time for $n_s \sim \xi t \geq 1$.  
The transition from a spontaneous to stimulated nature of the Stokes process also affects the spatial distribution of the atomic spin wave. 
For $n_s \le 1$ the excitation is calculated to be uniformly distributed 
in the  cell, while for $n_s > 1$ the spin-wave amplitude grows toward the end of the cell (see inset to Fig. 2b).  The observed dynamics 
provide evidence for the collective nature of the atomic spin excitations.  

After  a  time delay $\tau_{d}$, we apply the retrieve beam to convert the stored spin wave into anti-Stokes photons.  Fig.~2b demonstrates that the duration and peak flux of the anti-Stokes pulse can be controlled via the intensity of the retrieve laser.  The resonant retrieve laser converts the spin coherence into a dark-state polariton, and eventually into an anti-Stokes photon.
Note that the retrieve laser establishes an EIT configuration for the generated anti-Stokes field, so that the anti-Stokes light propagates unabsorbed through the cell.   In the ideal limit of perfect EIT and large optical depth,  the temporal shape of the anti-Stokes pulse is equivalent to the spatial shape of the atomic spin coherence, delayed by the time required to propagate out of the atomic cell at the group velocity $v_{g}(t) \propto |\Omega_R(t)|^2$~\cite{fleischhauer00}.
For larger (smaller) retrieve laser intensity, the excitation is released faster (slower), while the amplitude changes  in such a way that the total number of anti-Stokes photons is always equal to the number of spin-wave excitations.  
In practice, decay of the spin coherence during the delay time $\tau_{d}$ and finite optical depth flatten and broaden the anti-Stokes pulse, reducing the total number of anti-Stokes photons which can be retrieved within the coherence time of the atomic memory, as indicated by theoretical calculations (Fig. 2b) based on Ref.~\cite{fleischhauer00}.  The detailed comparison between theory and experiment in Fig. 2 suggests that the bandwidth of the generated anti-Stokes pulse is close to being  Fourier-transform limited, while the transverse profile effectively corresponds to only a few spatial modes.

At fixed laser intensities and durations, the number of 
Stokes and anti-Stokes photons fluctuates from event to event 
in a highly correlated manner~\cite{vdwal03etal}. In order to quantify these correlations, we directly compare  the number of Stokes and anti-Stokes photons for a large number of pulsed events (each with identical delay times $\tau_{d}$ and laser parameters).   The variance of the resulting distributions 
is then compared to the  photon shot noise level ${\rm PSN}_{th} = {\bar n}_S + {\bar n}_{AS}$, which represents the maximum degree of correlations possible for classical states~\cite{mandelbook}. In the experiment, great care is taken to eliminate systematic sources of errors, in particular 
APD dead-time effects.  To this end we experimentally determine photon shot noise for each channel by using a 50-50 beamsplitter and two APDs per detection channel (see Fig.~1c) which allows us to accurately determine the measured ${\rm PSN}_{meas} =  {\rm var}(AS1-AS2) + {\rm var}(S1-S2)$ value for each experiment. For correlation measurements  we typically choose the excitation intensities such that the average number of photons in each channel is on the order of or smaller than unity. Under such conditions the measured PSN approaches 
the expected, theoretical value of PSN.   To quantify the correlations, we consider the normalized variance $\mbox{V} = {\mbox{var}(\{n_{AS}-n_{S}\})}/{\mbox{PSN}_{meas}}$, which is one for classically correlated pulses and zero for pulses exhibiting perfect number correlations.  Using this method, we measure $V = 0.942 \pm 0.006$ for the data shown in Fig.~3 at delay time $\tau_{d} = 0$~\cite{deadtime}.
 
 Fig.~3 shows the normalized variance $V$
 as a function of storage time 
 $\tau_d$.  
Non-classical correlations ($V < 1$) between Stokes and anti-Stokes pulses are clearly observed for storage times up to a few microseconds. The time scale over which the correlations decay is determined by the coherence properties of the atomic spin-wave: nonclassical correlations are obtained only as long as the coherence of the stored excitation is preserved.   
Fig.~3 also shows that the retrieval efficiency (the ratio of the average number of anti-Stokes photons to the average number of Stokes photons) decreases in a similar manner as $\tau_d$ is increased. A fit to this time dependence (dotted line) yields a characteristic decoherence time $1/\gamma_c$ of about $3~\mu s$, consistent with the timescale for atomic diffusion from the write laser beam. 
These results demonstrate that within the spin coherence time, it is possible  to control the timing between preparation and retrieval, while preserving nonclassical correlations. 

\begin{figure}
\includegraphics[width=6.7cm]{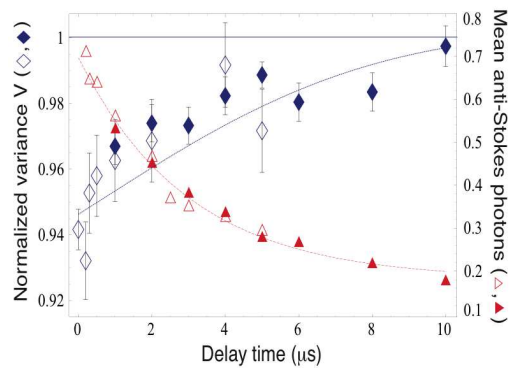}
\caption{\label{fig3}  
(Color Online) Observation of nonclassical correlations.  
Normalized variance V (blue) and mean number of anti-Stokes photons (red) versus delay time $\tau_{d}$.
The open and closed symbols represent two experimental runs with similar experimental parameters.
The dotted line is an exponential fit (characteristic time $\sim 3 \mu \mbox{s}$) to the mean number of anti-Stokes photons.  The solid line is the result of a theoretical model including the effects of loss, background, and several spatial modes on the Stokes and anti-Stokes channels~\cite{fig3model}.  
}
\end{figure}



It is important to note that at  $\tau_{d}=0$ the observed value  $V=0.942 \, \pm \, 0.006$ is far from the ideal value of
 $V=0$.    One source of error is the finite retrieval efficiency, which is limited by two factors.   Due to the atomic memory decoherence rate $\gamma_c$, the finite retrieval time $\tau_r$ always results in a finite loss probability $p \approx \gamma_c \tau_r$.  Moreover, even as $\gamma_c \rightarrow 0$ the retrieval efficiency is limited by the finite optical depth $\eta$ of the ensemble, which yields an error scaling as $p\sim 1/\sqrt{\eta}$.  
The anti-Stokes pulses in Fig. 3 have widths on the order of the measured decoherence time, so the atomic excitation decays before it is fully retrieved.    The measured maximum retrieval efficiency at $\tau_d =0$ corresponds to about 0.3.  In addition to finite retrieval efficiency, other factors reduce correlations, including losses in the detection system, background photons, APD afterpulsing effects, and imperfect mode matching. 

These correlations between Stokes and anti-Stokes pulses allow for the conditional preparation of the anti-Stokes pulse with intensity 
fluctuations that are suppressed  compared with classical light.   In order
to quantify the performance of  this technique, we measured the 
second-order intensity correlation function  $g^{(2)}_{n_{S}} (AS) $ and 
mean number of photons ${\bar n}^{AS}_{n_{S}}$
for the anti-Stokes pulse conditioned on the detection of $n_S$ photons in the Stokes channel (see Fig. 4).  (For classical states of light, $g^{(2)}\geq 1,$ whereas  an ideal  Fock state with $n$ photons has $g^{(2)} = 1-1/n$.)  Note that  the mean number of 
anti-Stokes photons grows linearly 
with $n_S$, while $g^{(2)}_{n_{S}} (AS) $ drops below unity, indicating the non-classical character of the anti-Stokes photon states. In the presence of background counts, $g^{(2)}_{n_{S}} (AS)$ does not increase monotonically with $n_S$, but instead exhibits a minimum at $n_S =2$.  The Mandel $Q$ parameter~\cite{mandelbook} can be calculated using  $Q^{AS}_{n_{S}}= {\bar n}^{AS}_{n_{S}} (g^{(2)}_{n_{S}} (AS) -1)$;  from the measurements we determine $Q^{AS}_{n_{S}=2} = -0.09 \ \pm \ 0.03$ for conditionally generated states with $n_{S} =2$
($Q \ge 0$ for classical states and $Q = -1$ for Fock states).  

\begin{figure}
\includegraphics[width=6.7cm]{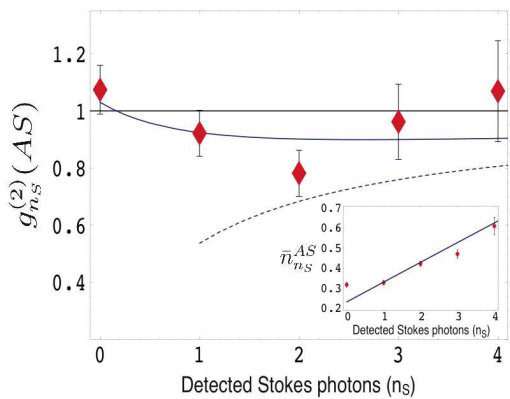}
\caption{\label{fig4} (Color Online) Conditional nonclassical state generation.   
Diamonds show experimentally measured values
$g^{(2)} (AS) = \langle AS_1 \cdot AS_2\rangle/\langle AS_1\rangle\langle AS_2\rangle$ (see Fig.~1c) as a function of the number of detected Stokes photons.  The measured mean photons numbers
were $\bar{n}_{S}=1.06$ and $\bar{n}_{AS}=0.36$.  The solid line shows the result of a theoretical model including background and loss on both the Stokes and anti-Stokes channels~\cite{fig4model}.  
({\bf Inset}) Measured mean anti-Stokes number  $\bar{n}^{AS}_{n_{S}}$ conditioned on the Stokes photon number $n_S$.  The solid line represents $\bar{n}^{AS}_{n_{S}}$  as predicted by the model.
}
\end{figure}

The observed reduction in the intensity fluctuations is imperfect 
due to background and losses in both the preparation and the retrieval detection channels. These can be accounted for in a theoretical model that yields reasonable agreement with experimental observations (solid curve in Fig. 4).  
Using this model corrected for loss and background on the retrieval channel (dotted line in Fig. 4), we estimate (${\bar n}^{AS}_{n_{S}=2}$, $Q^{AS}_{n_{S}=2}$) 
to be approximately (2.5, -0.85).
 
Although  the corrected parameters
are closer to the ideal limit, they still do not correspond to a perfect Fock state.   This is due to  loss and background in the preparation channel, which prevent measurement of the exact number of created spin excitations. In principle, the conditional  state preparation can 
be made insensitive to overall Stokes detection efficiency $\alpha_{S}$ by working in the regime of  a very weak excitation~\cite{duan01etal}; however, Stokes channel
background counts $n_{S}^{BG}$ prevent one from reaching this regime
in practice. A qualitative condition for high quality Fock state
generation, $\zeta \equiv n_{S}^{BG}(1-\alpha_{S}) / n_{S} \alpha_{S} \ll 1$, is only marginally fulfilled  in our experiments ($\zeta \sim 0.3$), accounting for the imperfectly prepared atomic states.   
Refinements in the Stokes detection system and better transverse mode selection should permit 
conditional Fock state generation with much greater purity, thereby providing a basic building block for 
long-distance quantum communication~\cite{duan01etal}.

We gratefully acknowledge  T.~Zibrova, J.~MacArthur, D.~Phillips, R.~Walsworth, P.~Hemmer, A.~Trifonov, and C. van der Wal for useful discussions and
experimental help. This  work was supported by the NSF, DARPA,
the 
Packard, 
Sloan, and 
Hertz Foundations, and \'{E}cole des Mines de Paris (FM).


\begin{thebibliography}{27}
\bibitem[{dea()}]{deadtime}
\bibinfo{note}{For the experimental conditions of Fig.~3, we estimate that
  dead-time effects reduce the measured $\bar{n}_{AS}$ by $2.5 \% $ from its
  actual value and increase $\mbox{V} = \mbox{var}(\{ n_{AS} - n_{S} \}) /
  \mbox{PSN}_{meas}$ by less than $0.2 \%$ from its actual value (well within
  the $\sim 0.6 \%$ error bars).}
\bibitem[{fig({\natexlab{a}})}]{fig3model}
\bibinfo{note}{Based on experimental measurements, the overall detection
  efficiency ($\alpha$) and number of background photons ($n_{BG}$) used in the
  model are $\alpha_{S} = 0.07$, $n^{BG}_{S} = 0.3$ ($\alpha_{AS} = 0.21$,
  $n^{BG}_{AS} = 0.12$) on the Stokes (anti-Stokes) channel, with a retrieval
  efficiency characteristic decay time of $3 \, \mu s$ and 4 transverse spatial
  modes assumed.}
\end{thebibliography}
\end{document}